**Title:** Using street view images and visual LLMs to predict heritage values for governance support: Risks, ethics, and policy implications

Tim Johansson[1], Mikael Mangold[1,2,*], Kristina Dabrock[3], Anna Donarelli[4], Ingrid Campo-Ruiz[1]

[1] RISE Research Institutes of Sweden AB, Built Environment, Gothenburg, Sweden

[2] Malmö University, Urban Studies, Malmö, Sweden

[3] Forschungszentrum Jülich GmbH, Institute of Climate and Energy Systems, Jülich Systems Analysis, 52425 Jülich, Germany

[4] Uppsala University, Department of Archaeology, Ancient History and Conservation, Uppsala, Sweden

*corresponding author: Mikael Mangold, mikael.mangold@ri.se

**Abstract:**

During 2025 and 2026, the Energy Performance of Buildings Directive is being implemented in the European Union member states, requiring all member states to have National Building Renovation Plans. In Sweden, there is no comprehensive national register of buildings with heritage values. This is seen as a barrier for the analyses underlying the development of Building Renovation Plans by the involved Swedish authorities. The purpose of this research was to assist Swedish authorities in developing information on heritage values in the Swedish building stock. Buildings in street view images from all over Sweden (N=154 710) have been analysed using multimodal Large Language Models (LLM) to assess visible aspects indicative of heritage value. Zero-shot predictions by LLMs were used as a basis for identifying buildings with potential heritage values for 5.0 million square meters of heated floor area. In this paper, the results of the predictions and lessons learned are presented and related to the development of the Swedish Building Renovation Plan as part of governance. The problems with the method and potential improvements are discussed. Risks with authorities' use of LLM-based data are addressed, with a focus on issues of transparency, error detection and sycophancy.

# 1 Introduction

Multimodal Large Language Models (LLM) can not only write text and produce images, but they can also be used to interpret and label texts and images, making large, complicated datasets understandable with new depth[1–3]. In the field of building stock analysis, street view imagery of buildings constitutes a complicated dataset that has previously been cumbersome to analyse[4–7]. LLMs in building stock analyses enable both research and decision support for authorities. However, when using the results of analyses based on LLMs, there are several uncertainties and pitfalls that need to be considered, such as the reliability of zero-shot classifications, the opacity of model decision-making processes, the potential for overconfidence and bias in generated outputs, and the juridical and ethical implications of applying such results in official decision-making.

Even though the technological developments are promising for creating a better overview of heritage values in the building stock, there are risks that authorities apply the tools without thorough scrutiny of the risks involved. For this reason, this paper examines a project in which Swedish authorities experimented with visual LLMs to predict heritage indicators for governance purposes.

The purpose of this paper is to describe and discuss a process where the Swedish National Heritage Board applied Large Language Models to assess building data. The study presents lessons learned from translating the outputs into policy and planning. The specific case concerns the European Union's Energy Performance of Buildings Directive (EPBD), which has the purpose of diminishing the energy use and climate impact from buildings. Buildings with heritage values may require special considerations in renovations but can and should be improved as well. The EPBD requires all member states to develop and update national building renovation plans every five years, starting in 2026. These plans must include an overview of the national building stock, a roadmap for energy improvements, a description of planned policy measures and an outline of investment and financing schemes. Data on buildings with heritage values can be included in order to guide possible exemptions or adapted energy requirements. Heritage value is the term used in Swedish legislation for what is considered valuable in the physical built environment from a historic, aesthetic, and social perspective. These three perspectives are both complementary and overlapping[8]. Heritage values are neither intrinsic nor universal; they are the results of assessments of both physical and intangible aspects in the built environment. Heritage is created in the present in order to represent history, and the selection of what is to be considered heritage, and having value as such, is based on contemporary needs and tastes[9]. There is heritage acknowledged and officially protected by international, national and local institutions, and there is also unofficial heritage which can be of great importance to specific groups or communities[10]. Character-defining elements are material manifestations in buildings that make it possible to understand their context and that therefore may add to their heritage value[11]. Such elements can be visibly detected and constitute one of several aspects used in the assessment of heritage values. In Sweden, there is no comprehensive national register of buildings with heritage values[12].

Protection of buildings with heritage values is stipulated in the Historic Environment Act[13] (HEA) and in the Planning and Building Act[14] (PBA). Buildings can be protected through official listings according to the HEA, and such buildings are managed at the national or regional level. The PBA protects heritage values in buildings through the "prohibition of distortion" and the "requirement of caution". The prohibition of distortion applies to buildings that are considered particularly valuable from a historic point of view, whereas the requirement of caution applies to all existing buildings. The aim is to preserve characteristics and qualities in the existing building stock. It can be determined during, for instance, a building permit process that a building qualifies as particularly valuable, it does not have to be listed in advance, but several municipalities try to keep records of them.

Existing records of buildings with heritage values therefore consist of listed buildings protected by the HEA and, to some extent, municipal lists of buildings protected by the PBA. These assessments have been done at different times in history using information compiled through different methods, including but not limited to ocular surveys of exteriors and/or interiors, archival searches and different material analyses. Buildings and the reasons for their designation are thereby also described with varying degrees of depth in these registers. For the sake of the EPBD implementation, the Swedish National Heritage Board found they needed to complete and update these existing records, and that novel tools for deriving building stock information could prove efficient for this purpose. Specifically, the lack of data on buildings protected by the PBA is considered a problem. By using LLM, the authority hoped to fill in gaps and improve the knowledge of the building stock.

The varying information necessary to assess heritage values can make it difficult to determine what kind of information is useful in different policy situations. The LLM assessments were limited to visible features in street view images that could be indicative of heritage values in buildings. The visible façade elements that can be detected in street view images constitute a small part of what needs to be considered to thoroughly assess a building's heritage value. Another limitation with the LLM is that the façade is not seen in its environmental context, but each façade is assessed individually. It can only be used to get an idea of a specific building's characteristics. Surveys of built environments are being done by heritage experts in different regions as part of characterising built environments for the purpose of managing heritage values[15]. It is important to stress that these physical surveys, in which experts assess facades and put them in the context of the whole environment, are only a first step of a complete assessment method. Data such as the history and use of the place will need to be added before any evaluations or plans for development can be made.

The authors of this paper have previously worked as researchers for the Swedish Board of Housing, Building and Planning (Boverket) and the Swedish National Heritage Board (Riksantikvarieämbetet) in several research tasks, producing decision support documents for the building sector[16]. Methods for analysing register information have been developed over many years and have supported the process described in this paper[17–19]. During the winter of 2024/2025, the authors of this paper prompted OpenAI's GPT (gpt-4o-2024-08-06) 281 081 times using the prompt found in Supplementary Material 1 and attached street view images of the buildings. As several buildings were associated with multiple camera viewpoints, and some of the camera points had lower visibility, this corresponded to a total of 154,710 unique Energy Performance Certificates (EPC) that were successfully assessed. Not all buildings could be evaluated, for instance, due to obstructions in the images or other limitations[20].

The research field of applying LLMs in assessing aspects of heritage values is novel, and the few examples that exist have been published in recent years[21,22]. The inclusion of AI in contemporary cartography proposes to view mapping as a network of relationships between human and machine intelligences[23]. The automation of some tasks with LLMs can be positive when supporting scarce human resources in urban planning tasks[24]. For assessing architectural typologies, LLMs can facilitate access to plural stakeholder perspectives beyond traditional analytical methods[25]. For building stock in particular, attributes like building height, construction year and type have been collected across 27 European Union countries and Switzerland using government datasets and OpenStreetMap[26]. Assessing of LLM-based built environment audits shows that while their results are highly accurate, the possible errors could lead to false evaluations unless there are human reviews in the process[27]. However, the use of LLMs in urban climate change mitigation remains negligible, probably due to data availability and commercial interest[28]. LLMs have been analysed to assess intangible heritage, results showing that their linguistic capability affected their performance, their ability to understand and use information varied, and that hallucinations were reduced when context was provided[29]. Researchers have found that bias against specific regions or cultures is the hardest to fix, requiring expert input from heritage specialists[30]. Building stocks are nation-specific, and in many cases even regionally specific. Historical events, legislation, available natural resources and traditions are all highly specific and affect how the building stock has been built and developed.

At the core of this study is the use of zero-shot prompting, where the LLM is asked to rank the heritage value indicators of buildings, based on what is visible in a façade image, directly without additional training data or demonstrations. Zero-shot prompting means that the prompt used to interact with the LLM does not contain examples or demonstrations. The zero-shot prompt directly asks the LLM for the variable of interest without any additional examples to steer it. This is a rather crude way of using AI, but it has some advantages:

- It is faster if the training of models can be avoided, since generating training data is time-consuming.
- It is easier to communicate the generated results to domain experts and authorities who are less familiar with statistics and machine learning.
- The domain experts' own tools can be applied directly in the prompt. In this case, the authorities' checklists for assigning heritage values were used to build the prompt[31].

Furthermore, the quality of LLMs' answers has continuously increased and has been shown to be competitive with fine-tuned machine learning models for various building attributes[32].

However, the usage of AI in governance raises questions regarding reliability, transparency, and appropriateness. LLMs are often overconfident, prone to sycophancy, and trained to be well-spoken; human beings are easily impressed by such traits[33]. The human species even evolves based on traits of being well-spoken. Arguably, this is one reason why people are so impressed by the capabilities of LLMs. In addition to these methodological and cognitive challenges, the application of LLM-based analyses in governance also intersects with broader institutional, legal and ethical considerations.

The Swedish government has called for increased use of AI in public authorities to enhance efficiency, while at the same time, there is hesitancy and uncertainty regarding its role in decision-making[34,35]. Further complications arise from the fact that both LLMs and street view databases are owned by companies outside Sweden and the EU. The images used were extracted from Google Street View, whose terms restrict use, while alternatives such as Mapillary offer broader rights but face quality and coverage limitations[36]. These uncertainties also extend to the legal framework: although the EU's copyright directive (Article 3) allows data mining for academic use, its implementation has yet to be tested in court[37]. These issues have been raised with national and European authorities in the EPBD implementation process.

Against this background, this paper investigates the potentials and risks of using LLM-derived labels indicating heritage values for decision support in building stock governance in the particular case of the Swedish EPBD implementation, answering the following research questions:

1. How can authorities be assisted using information derived from zero-shot prompting and street view imagery?
2. What risks are there for authorities when using answers from zero-shot prompting in decision making?

The technical validation of the method is reported in a separate article[38]. This paper instead focuses on the governance implications, highlighting lessons learned and considerations for the future use of AI-derived labels in policy and planning.

# 2 Methods

The overall method is illustrated in Figure 1. The initial data processing and integration follow the procedures described by Dabrock et al.[38]. Briefly, the analysis builds on EPCs for 178,545 multifamily and 93,629 non-residential buildings, spatially linked to the national property register of approximately nine million buildings. Universally Unique Identifiers (UUIDs) and coordinates were used to enable georeferencing, metadata collection, and the calculation of sight lines. Building footprints (2D polygons) and road centerlines (polylines) were obtained from Swedish Land Survey (Lantmäteriet), providing the geometric basis for spatial joins and façade image generation.

**Overview of methodological steps**

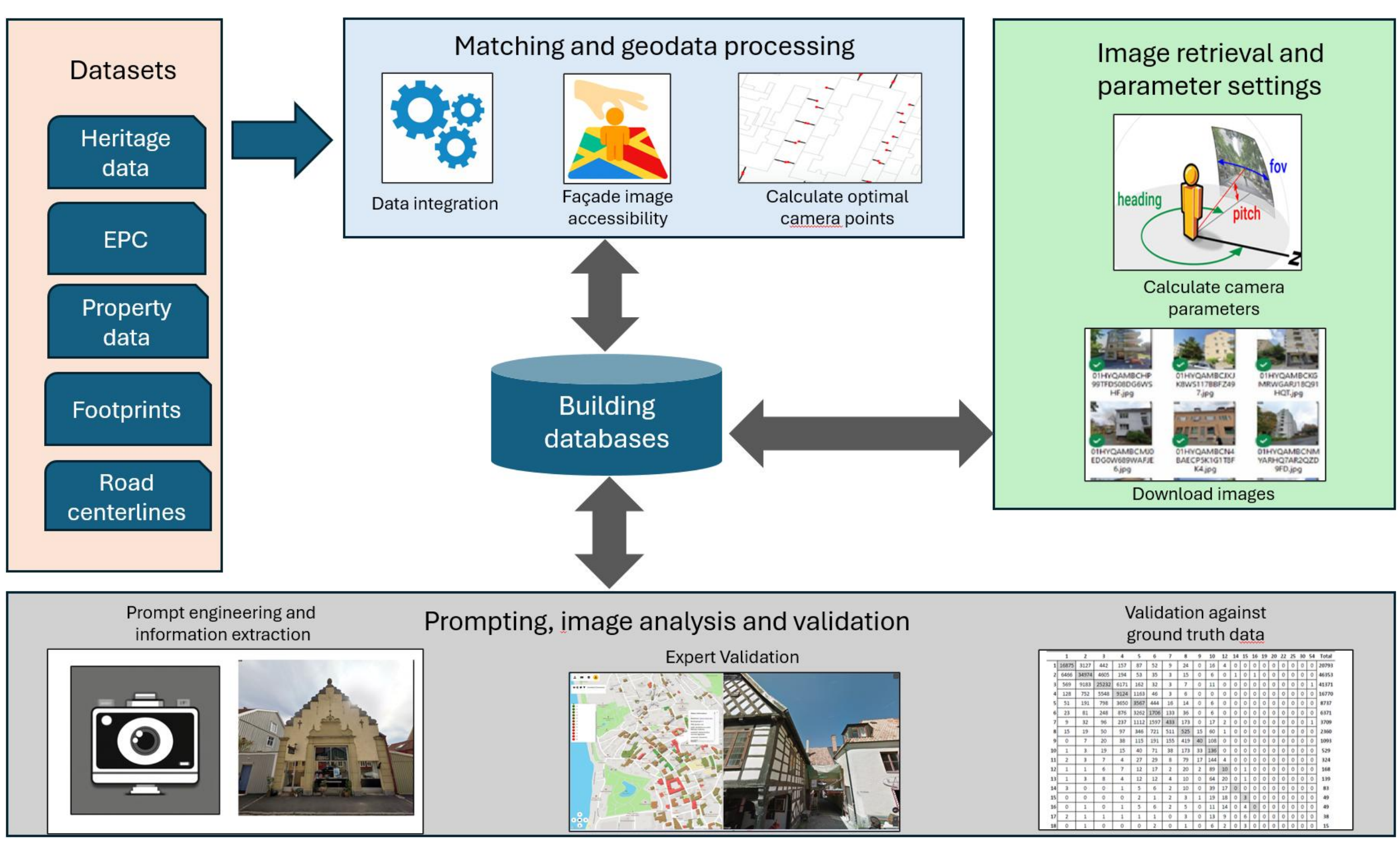

Fig. 1 Datasets list the sources of building and image data from which features are selected. Matching and Geodata Processing explains the linking of EPC records with the property register and associated geospatial processing. Image retrieval and parameter settings cover the acquisition of Street View images and the calculation of image parameters, while Prompting, Image Analysis and Validation detail the labelling of images using LLM and the validation procedures.

While EPCs provide large-scale coverage, the information on heritage value in the EPCs is unreliable[12]. Misclassifications are common, with officially protected heritage buildings remaining unrecognised. To establish a more reliable ground truth, we used the Database of Built Heritage, which includes over 80,000 buildings, of which around 13,000 are officially protected at the national level, including churches. The register is only complete with respect to these approximately 13,000 listed buildings protected by the HEA. They correspond to approximately 0.15% of the Swedish building stock. The remaining entries in the Database of Built Heritage are mainly buildings with heritage values identified through local or regional building surveys, and coverage varies considerably across different parts of the country.

As a result, a substantial share of buildings that would potentially be considered to represent heritage values remain absent from the dataset. To address this gap, we incorporated additional sources beyond the buildings officially protected at the national level. The "Räkna Q" dataset, which documents buildings protected under the PBA, was used to identify buildings officially protected at the regional and local levels. This dataset is compiled from detailed plans and area regulations available since 1987 in the archives of the County Administrative Boards, with coverage currently available for 19 of Sweden's 21 counties. It serves both for monitoring and environmental policy evaluation, while providing an independent benchmark for validating our method.

## 2.1 Matching and geodata processing

Building identifiers were derived by linking EPC records with the national property register. Buildings officially protected at the national level, Byggnadsminnen, could be directly integrated, while the Räkna Q dataset was matched to building footprints through spatial joins. Street-level imagery was then assessed using the Google Street View Metadata API. Coverage was extensive, with 86.6% of all buildings having images available within 100 meters. Approximately 87.3% of these images were captured between 2019 and 2024, providing sufficient temporal resolution for façade analysis. No images were excluded on the basis of age, as older imagery was considered preferable to having no visual data at all. To determine optimal camera positions, building footprints were segmented into wall sections and connected to the nearest road centreline from the national property map. Sightlines were filtered to exclude obstructed perspectives, and the maximum distance was adjusted from 100 meters to 50 meters in order to ensure higher-quality images. Additionally, sightlines with deviations exceeding three degrees from perpendicular were excluded. For each building, the shortest valid sightline was retained. Ultimately, 72% of all buildings had at least one valid camera position, corresponding to 281,018 usable viewpoints. The filtering of buildings is summarised in Table 1, which shows the reduction from all EPC buildings to the subset for which, after matching and image availability checks, a usable Google Street View image could be retrieved and used for observing a building.

Table 1 Buildings, identifiers and availability of street view images for the analyses.

| Category | EPC buildings | Buildings with at least one usable image | Observations of buildings |
|---|---|---|---|
| Multifamily | 178,545 | 127 333 | 107 799 |
| Non-residential | 93,629 | 58 705 | 46 911 |
| Total | 272,174 | 186 038 | 154 710 |

## 2.2 Image retrieval and parameter settings

In this step, the camera pitch was derived from estimated building heights, obtained from the EPCs by multiplying the reported number of floors by 3.0 meters, combined with the distance between the façade and the camera location[7]. The zoom parameter was defined as a function of wall width and the distance to the photo point, ensuring consistent framing of the façade across different building sizes. Finally, façade images were retrieved from the Google Street View API, using geographic coordinates together with the calculated heading, pitch, and zoom as input parameters. Images were retrieved regionally, with separate processing for multifamily and non-residential buildings to ensure systematic coverage.

## 2.3 Prompting and image analysis

In this step, façade images served as the primary input for information extraction using OpenAI's GPT (gpt-4o-2024-08-06). In total, the model was prompted with 281,018 images representing 186,038 buildings. Buildings with a model-assigned visibility score below 50 were excluded, resulting in 179,538 observed buildings, related to 154,710 EPCs. For EPCs with multiple images, the building with the highest predicted heritage indicator score was selected to represent the EPC. All subsequent analyses were conducted at the EPC level, counting each certificate once and aggregating the heated floor area within the same certificate. This avoids double-counting, as buildings sharing a certificate typically also share many building and energy-related attributes, making them statistically non-independent observations.

The initial approach of directly querying the model to classify buildings into high or low heritage categories was abandoned due to inconsistent outputs. Instead, a more structured prompting strategy was adopted, building on recent insights that longer and more detailed prompts generally improve model performance[39]. The refined strategy applied the Question Refinement Pattern[40], in which the model was asked to assist in formulating the question itself.

Following the approach of Ekin[41], the LLM was assigned a persona and instructed to complete a predefined template with specified attributes. To ensure consistency and facilitate subsequent analysis, the responses were required in JSON format. The prompt design was informed by multiple sources: the draft checklist for ocular building surveys developed by the Swedish National Heritage Board, Boverket's[31] official checklist, and the feature typology described by Björk et al.[42] for Swedish multi-family housing. Input from workshops with the Swedish National Heritage Board and Uppsala University further contributed to refining the prompt. The final prompt structure and selected features are presented in Supplementary Material 1.

## 2.4 Expert validation

To facilitate interpretation and exploration, a web-based mapping tool was developed, integrating Mapbox and Google Street View. As shown in Figure 2, the platform presents each building's façade image together with extracted features, predicted heritage indicator, and register-based attributes in a unified interface, supporting both expert review and further analysis.

In December 2024, structured workshops were held at Gotland with heritage experts from the Swedish National Heritage Board and Uppsala University. Both the EU, through the AI Act[43], and the UK, through its Generative AI Framework[44], emphasise human oversight as a key measure to mitigate errors and risks. A common method for achieving this is to implement a secondary manual quality assurance step, in which a qualified human expert reviews and, if necessary, corrects the material before it is put to use[45]. During the workshops, validations were conducted, and decisions in regard to its use in the implementation of the EPBD were taken. Importantly, expert validation was fitted to the intended implementation. It was decided to use the predicted heritage indicators only on a building stock level for this specific purpose. For other researchers who develop AI tools and analyses for authorities to be used in decision making, we stress the importance of adapting tools, analyses and validation to the specific user case. For this reason, tools need to be flexible, interpretable and allow for rapidly answering to the domain expert and decision makers' perspectives.

**Visualisation tool for showing results to experts at the authorities**

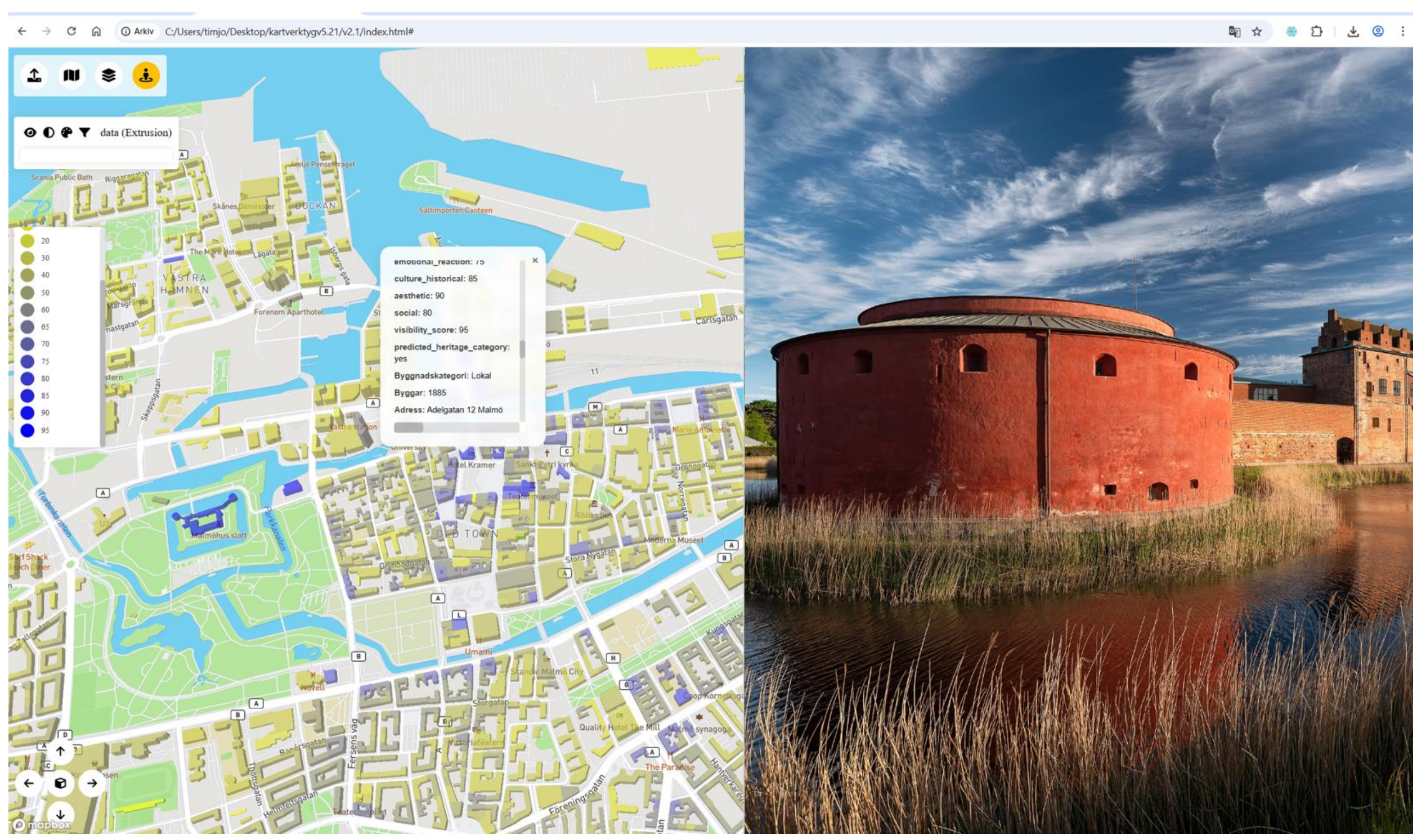

Fig. 2 The web-based tool, with the 3D map coloured by predicted heritage indicator (yellow to blue) and the corresponding image (Image credits: Kateryna Baiduzha, CC BY-SA 4.0). For copyright reasons, the Google street view building image has been replaced with an open licence street view image.

# 3 Results

The results chapter first compares the GPT predicted parameters with available Ground truth features. However, the existing records of buildings with heritage values are complex and not as easily applicable as ground truth features. In the second section, the predicted heritage indicators are compared to the national and local registers of officially protected heritage buildings (the Database of Built Heritage and Räkna Q). In the third section, the bias and representativeness of the predicted heritage indicators are analysed using auxiliary building stock information. The final results chapter focuses on the process of using the results in the EPBD implementation in Sweden.

## 3.1 Cross-Data Comparison with Auxiliary Data

In the prompt used in December 2024 (see Supplementary Material 1), building age and number of floors were asked for; these variables are also included in the Energy Performance Certificate. A cross-comparison between the predicted features and the variables in the registers makes it possible to discern patterns in the predictions, see Tables 2 and 3. On the building specific level, there are discrepancies between the predicted values and the register data, but on the aggregated levels, there is some alignment with 'Ground truth' in register data. The predicted number of floors in Table 3 is more well-aligned. The building that has 54 floors in Table 3 is Turning Torso in Malmö. Turning Torso is well documented in online texts, and OpenAI's GPT has correctly identified it in the images. Turning Torso was also visible in the background behind some buildings that had been torn down and have consequently been identified two additional times. Another observation is that OpenAI's GPT never predicted a building to have 13 floors.

In Table 2, some patterns can be seen in the predictions by OpenAI's GPT. Few buildings were assigned construction years in the 1910s, 1940s, 1990s and 2010s. This could be the result of (un)educated guesses that fewer buildings were built during the World War periods and the property and bank crises. OpenAI's GPT have also overestimated the number of buildings that are a 1000 years old.

Table 2 Predicted construction periods (horizontally) compared to construction periods in the Energy Performance Certificates (vertically). The unit is an Energy Performance Certificate/prediction. Grey cells are those where predicted values match the Energy Performance Certificate construction period.

| | **Predicted construction periods** | | | | | | | | | | | | | | | | | | | | | |
|---|---|---|---|---|---|---|---|---|---|---|---|---|---|---|---|---|---|---|---|---|---|
| **Construction periods in the Energy Performance Certificates** | **1000** | **1200** | **1300** | **1400** | **1500** | **1600** | **1700** | **1800** | **1900** | **1910** | **1920** | **1930** | **1940** | **1950** | **1960** | **1970** | **1980** | **1990** | **2000** | **2010** | **2020** | **Total** |
| **1000** | 0 | 0 | 0 | 0 | 0 | 0 | 0 | 0 | 2 | 0 | 0 | 0 | 0 | 2 | 0 | 2 | 0 | 0 | 0 | 0 | 0 | **6** |
| **1100** | 0 | 0 | 1 | 0 | 0 | 0 | 0 | 0 | 0 | 0 | 0 | 0 | 0 | 1 | 1 | 1 | 1 | 0 | 0 | 0 | 0 | **5** |
| **1200** | 0 | 1 | 3 | 0 | 1 | 0 | 0 | 5 | 0 | 0 | 0 | 0 | 0 | 0 | 0 | 0 | 0 | 0 | 0 | 0 | 0 | **10** |
| **1300** | 0 | 0 | 0 | 0 | 1 | 2 | 0 | 5 | 1 | 0 | 1 | 0 | 0 | 0 | 0 | 0 | 0 | 0 | 0 | 0 | 0 | **10** |
| **1400** | 0 | 1 | 1 | 0 | 1 | 1 | 1 | 7 | 3 | 0 | 1 | 0 | 0 | 0 | 0 | 0 | 0 | 0 | 0 | 0 | 0 | **16** |
| **1500** | 0 | 0 | 1 | 0 | 0 | 2 | 4 | 7 | 1 | 0 | 1 | 0 | 0 | 0 | 0 | 0 | 0 | 0 | 0 | 0 | 0 | **16** |
| **1600** | 12 | 0 | 1 | 1 | 0 | 18 | 45 | 104 | 60 | 0 | 5 | 0 | 0 | 3 | 2 | 1 | 1 | 0 | 0 | 0 | 2 | **255** |
| **1700** | 29 | 1 | 2 | 0 | 1 | 11 | 36 | 337 | 360 | 0 | 12 | 1 | 0 | 13 | 0 | 3 | 1 | 0 | 1 | 0 | 5 | **813** |
| **1800** | 56 | 2 | 0 | 0 | 3 | 6 | 8 | 711 | 3307 | 4 | 234 | 0 | 1 | 276 | 8 | 25 | 13 | 0 | 36 | 1 | 36 | **4727** |
| **1900** | 26 | 1 | 0 | 0 | 1 | 2 | 6 | 113 | 2511 | 3 | 576 | 4 | 0 | 465 | 13 | 54 | 24 | 2 | 57 | 0 | 31 | **3889** |
| **1910** | 8 | 0 | 0 | 0 | 1 | 0 | 1 | 42 | 1015 | 34 | 1022 | 15 | 3 | 351 | 5 | 36 | 10 | 2 | 36 | 0 | 19 | **2600** |
| **1920** | 42 | 1 | 0 | 0 | 0 | 2 | 16 | 316 | 4094 | 26 | 2667 | 100 | 14 | 2150 | 47 | 196 | 70 | 1 | 116 | 1 | 76 | **9935** |
| **1930** | 28 | 0 | 0 | 0 | 3 | 1 | 3 | 53 | 1034 | 2 | 1259 | 288 | 25 | 7596 | 312 | 551 | 65 | 1 | 148 | 0 | 80 | **11449** |
| **1940** | 28 | 0 | 0 | 0 | 0 | 0 | 2 | 26 | 453 | 1 | 252 | 21 | 23 | 11575 | 1181 | 1497 | 111 | 2 | 269 | 0 | 156 | **15597** |
| **1950** | 47 | 0 | 0 | 0 | 0 | 1 | 0 | 35 | 420 | 1 | 171 | 14 | 5 | 10584 | 3034 | 3518 | 190 | 3 | 435 | 0 | 258 | **18716** |
| **1960** | 127 | 1 | 0 | 0 | 0 | 1 | 0 | 20 | 245 | 0 | 81 | 1 | 0 | 4291 | 4489 | 10525 | 549 | 11 | 1330 | 2 | 878 | **22551** |
| **1970** | 135 | 1 | 0 | 0 | 1 | 2 | 0 | 18 | 231 | 0 | 68 | 1 | 2 | 1105 | 1075 | 7854 | 1099 | 14 | 2075 | 1 | 1106 | **14788** |
| **1980** | 121 | 0 | 1 | 0 | 0 | 0 | 0 | 30 | 607 | 0 | 207 | 17 | 2 | 1766 | 428 | 5651 | 3077 | 110 | 2782 | 2 | 1109 | **15910** |
| **1990** | 71 | 0 | 0 | 0 | 0 | 0 | 0 | 22 | 417 | 1 | 210 | 33 | 1 | 1885 | 324 | 3319 | 3011 | 347 | 3136 | 0 | 852 | **13629** |
| **2000** | 44 | 0 | 0 | 0 | 0 | 0 | 0 | 8 | 160 | 0 | 70 | 5 | 0 | 758 | 184 | 1114 | 734 | 111 | 4188 | 15 | 1317 | **8708** |
| **2010** | 32 | 0 | 0 | 0 | 0 | 0 | 0 | 4 | 106 | 0 | 44 | 1 | 1 | 331 | 123 | 800 | 266 | 21 | 3585 | 134 | 3253 | **8701** |
| **2020** | 6 | 0 | 0 | 0 | 0 | 0 | 0 | 2 | 23 | 0 | 10 | 0 | 0 | 77 | 21 | 138 | 28 | 1 | 452 | 8 | 1607 | **2373** |
| **Total** | **812** | **9** | **10** | **1** | **13** | **49** | **122** | **1865** | **15050** | **72** | **6891** | **501** | **77** | **43229** | **11247** | **35285** | **9250** | **626** | **18646** | **164** | **10785** | **154710** |

Table 3 Predicted number of floors (horizontally) compared to floors in the Energy Performance Certificates (vertically). The unit is an Energy Performance Certificate/prediction. Grey cells are those where predicted values match the number of floors registered in the Energy Performance Certificates.

| Floors in the Energy Performance Certificates | Predicted number of floors | | | | | | | | | | | | | | | | | | | | |
|---|---|---|---|---|---|---|---|---|---|---|---|---|---|---|---|---|---|---|---|---|---|
| | 1 | 2 | 3 | 4 | 5 | 6 | 7 | 8 | 9 | 10 | 12 | 14 | 15 | 16 | 19 | 20 | 22 | 25 | 30 | 54 | Total |
| 1 | 16875 | 3127 | 442 | 157 | 87 | 52 | 9 | 24 | 0 | 16 | 4 | 0 | 0 | 0 | 0 | 0 | 0 | 0 | 0 | 0 | 20793 |
| 2 | 6466 | 34974 | 4605 | 194 | 53 | 35 | 3 | 15 | 0 | 6 | 0 | 1 | 0 | 1 | 0 | 0 | 0 | 0 | 0 | 0 | 46353 |
| 3 | 569 | 9183 | 25232 | 6171 | 162 | 32 | 3 | 7 | 0 | 11 | 0 | 0 | 0 | 0 | 0 | 0 | 0 | 0 | 0 | 1 | 41371 |
| 4 | 128 | 752 | 5548 | 9124 | 1163 | 46 | 3 | 6 | 0 | 0 | 0 | 0 | 0 | 0 | 0 | 0 | 0 | 0 | 0 | 0 | 16770 |
| 5 | 51 | 191 | 798 | 3650 | 3567 | 444 | 16 | 14 | 0 | 6 | 0 | 0 | 0 | 0 | 0 | 0 | 0 | 0 | 0 | 0 | 8737 |
| 6 | 23 | 81 | 248 | 876 | 3262 | 1706 | 133 | 36 | 0 | 6 | 0 | 0 | 0 | 0 | 0 | 0 | 0 | 0 | 0 | 0 | 6371 |
| 7 | 9 | 32 | 96 | 237 | 1112 | 1597 | 433 | 173 | 0 | 17 | 2 | 0 | 0 | 0 | 0 | 0 | 0 | 0 | 0 | 1 | 3709 |
| 8 | 15 | 19 | 50 | 97 | 346 | 721 | 511 | 525 | 15 | 60 | 1 | 0 | 0 | 0 | 0 | 0 | 0 | 0 | 0 | 0 | 2360 |
| 9 | 0 | 7 | 20 | 38 | 115 | 191 | 155 | 419 | 40 | 108 | 0 | 0 | 0 | 0 | 0 | 0 | 0 | 0 | 0 | 0 | 1093 |
| 10 | 1 | 3 | 19 | 15 | 40 | 71 | 38 | 173 | 33 | 136 | 0 | 0 | 0 | 0 | 0 | 0 | 0 | 0 | 0 | 0 | 529 |
| 11 | 2 | 3 | 7 | 4 | 27 | 29 | 8 | 79 | 17 | 144 | 4 | 0 | 0 | 0 | 0 | 0 | 0 | 0 | 0 | 0 | 324 |
| 12 | 1 | 1 | 6 | 7 | 12 | 17 | 2 | 20 | 2 | 89 | 10 | 0 | 1 | 0 | 0 | 0 | 0 | 0 | 0 | 0 | 168 |
| 13 | 1 | 3 | 8 | 4 | 12 | 12 | 4 | 10 | 0 | 64 | 20 | 0 | 1 | 0 | 0 | 0 | 0 | 0 | 0 | 0 | 139 |
| 14 | 3 | 0 | 0 | 1 | 5 | 6 | 2 | 10 | 0 | 39 | 17 | 0 | 0 | 0 | 0 | 0 | 0 | 0 | 0 | 0 | 83 |
| 15 | 0 | 0 | 0 | 0 | 2 | 1 | 2 | 3 | 1 | 19 | 18 | 0 | 3 | 0 | 0 | 0 | 0 | 0 | 0 | 0 | 49 |
| 16 | 0 | 1 | 0 | 1 | 5 | 6 | 2 | 5 | 0 | 11 | 14 | 0 | 4 | 0 | 0 | 0 | 0 | 0 | 0 | 0 | 49 |
| 17 | 2 | 1 | 1 | 1 | 1 | 1 | 0 | 3 | 0 | 13 | 9 | 0 | 6 | 0 | 0 | 0 | 0 | 0 | 0 | 0 | 38 |
| 18 | 0 | 1 | 0 | 0 | 0 | 2 | 0 | 1 | 0 | 6 | 2 | 0 | 3 | 0 | 0 | 0 | 0 | 0 | 0 | 0 | 15 |
| 19 | 0 | 0 | 0 | 0 | 0 | 0 | 0 | 1 | 0 | 2 | 1 | 0 | 3 | 0 | 1 | 0 | 0 | 0 | 0 | 0 | 8 |
| 20 | 0 | 1 | 1 | 0 | 0 | 2 | 0 | 0 | 0 | 1 | 0 | 0 | 1 | 0 | 0 | 2 | 0 | 0 | 0 | 0 | 8 |
| 21 | 0 | 0 | 1 | 0 | 1 | 0 | 0 | 0 | 0 | 0 | 0 | 0 | 2 | 0 | 0 | 0 | 0 | 0 | 0 | 0 | 4 |
| 22 | 1 | 0 | 0 | 0 | 0 | 0 | 0 | 0 | 0 | 3 | 2 | 0 | 2 | 0 | 0 | 0 | 1 | 0 | 0 | 0 | 9 |
| 23 | 0 | 0 | 0 | 0 | 1 | 0 | 0 | 0 | 0 | 0 | 0 | 0 | 1 | 0 | 0 | 0 | 0 | 0 | 0 | 0 | 2 |
| 24 | 0 | 0 | 0 | 0 | 1 | 1 | 0 | 0 | 0 | 0 | 0 | 0 | 0 | 0 | 0 | 0 | 0 | 0 | 0 | 0 | 2 |
| 25 | 0 | 0 | 1 | 0 | 1 | 0 | 0 | 0 | 0 | 0 | 0 | 0 | 1 | 0 | 0 | 0 | 0 | 1 | 0 | 0 | 4 |
| 26 | 0 | 0 | 0 | 1 | 0 | 1 | 0 | 1 | 0 | 0 | 1 | 0 | 1 | 0 | 0 | 0 | 0 | 0 | 0 | 0 | 5 |
| 27 | 0 | 0 | 0 | 0 | 0 | 0 | 0 | 0 | 0 | 0 | 0 | 0 | 0 | 0 | 0 | 0 | 0 | 1 | 0 | 0 | 1 |
| 28 | 0 | 0 | 1 | 0 | 0 | 0 | 0 | 0 | 0 | 1 | 0 | 0 | 0 | 0 | 0 | 0 | 0 | 0 | 0 | 0 | 2 |
| 29 | 0 | 0 | 0 | 0 | 0 | 0 | 0 | 0 | 0 | 1 | 0 | 0 | 0 | 0 | 0 | 0 | 0 | 0 | 0 | 0 | 1 |
| 30 | 0 | 0 | 0 | 0 | 0 | 0 | 0 | 0 | 0 | 1 | 0 | 0 | 0 | 0 | 0 | 0 | 0 | 0 | 0 | 0 | 1 |
| 32 | 0 | 0 | 0 | 0 | 0 | 0 | 0 | 0 | 0 | 1 | 0 | 0 | 0 | 0 | 0 | 0 | 0 | 0 | 0 | 0 | 1 |
| 33 | 0 | 0 | 1 | 1 | 0 | 0 | 0 | 0 | 0 | 0 | 0 | 0 | 0 | 0 | 0 | 0 | 0 | 0 | 0 | 0 | 2 |
| 35 | 0 | 0 | 1 | 0 | 0 | 0 | 0 | 0 | 0 | 0 | 0 | 0 | 0 | 0 | 0 | 0 | 0 | 0 | 0 | 0 | 1 |
| 36 | 0 | 1 | 0 | 0 | 0 | 0 | 0 | 0 | 0 | 0 | 0 | 0 | 0 | 0 | 0 | 0 | 0 | 0 | 1 | 0 | 2 |
| 37 | 0 | 0 | 0 | 0 | 0 | 0 | 0 | 0 | 0 | 0 | 0 | 0 | 0 | 0 | 0 | 1 | 0 | 0 | 1 | 0 | 2 |
| 54 | 0 | 0 | 0 | 0 | 0 | 0 | 0 | 0 | 0 | 0 | 0 | 0 | 0 | 0 | 0 | 0 | 0 | 0 | 0 | 1 | 1 |
| tot | 24147 | 48381 | 37086 | 20579 | 9975 | 4973 | 1324 | 1525 | 108 | 761 | 105 | 1 | 29 | 1 | 1 | 3 | 1 | 2 | 2 | 3 | |

The prompt also included several aspects that are not registered in any central building register in Sweden. For these features, it is not possible to validate the predictions. In Table 4, Number of panes in the windows exemplifies one such feature. Number of panes in the windows also exemplifies one of the misunderstandings that resulted from the prompt. The number of panes has high relevance for the insulation properties of windows, and windows are considered an important character-defining element in buildings. Initial tests showed that LLMs could zero-shot identify number of panes based on reflections in windows. However, when all street view images were used, and the images had a larger distance between the camera and building, OpenAI's GPT instead interpreted the task to be the numbering of subdivisions with window bars. LLMs also have a limited capacity to count objects[46].

Table 4 Number of panes in the windows predicted for buildings in the heritage groups. The columns sum to 100%.

| Number of panes in the windows | Byggnads-minne | Räkna Q | Heritage value from age | Predicted heritage value | No assigned heritage value | Total |
|---|---|---|---|---|---|---|
| 0 | 0.03% | 0.06% | 0.08% | 0.00% | 0.28% | 0.25% |
| 1 | 13.95% | 26.12% | 10.69% | 22.71% | 58.91% | 53.58% |
| 2 | 20.24% | 32.94% | 42.40% | 24.37% | 36.209% | 36.29% |
| 3 | 0.00% | 0.11% | 0.09% | 0.42% | 0.14% | 0.14% |
| 4 | 17.00% | 13.19% | 15.41% | 14.97% | 2.95% | 4.42% |
| 6 | 45.31% | 25.56% | 29.80% | 35.99% | 1.46% | 5.04% |
| 8 | 1.98% | 1.26% | 1.16% | 1.31% | 0.04% | 0.19% |
| 9 | 0.00% | 0.00% | 0.00% | 0.00% | 0.00% | 0.00% |
| 10 | 0.10% | 0.00% | 0.01% | 0.00% | 0.00% | 0.00% |
| 12 | 1.37% | 0.44% | 0.30% | 0.21% | 0.01% | 0.05% |
| 20 | 0.00% | 0.27% | 0.01% | 0.00% | 0.00% | 0.01% |
| **Total [$Mm^2$]** | **2.95** | **7.61** | **26.60** | **5.00** | **304.09** | **346.25** |

## 3.2 The LLM-predicted heritage indicator

The prompt, in combination with the street-level image, led OpenAI's GPT to generate a large number of building-related parameters in response. Dabrock et al.[38] combined these parameters with registry data to train a separate AI model to predict heritage categories of buildings. While this method yielded good results in Stockholm, it could not be applied to the rest of Sweden due to the lack of a comprehensive and representative dataset covering all regions of the country for training and predicting heritage categories. This is due to the fact that different municipalities and regions lack comprehensive data, and that there are many local variations for such categories.

The Swedish National Heritage Board decided to instead directly use the predicted heritage indicator (1-100) numeric parameter to include additional buildings in the EPBD process. They defined an appropriate threshold high enough (50 or higher) to include 1% of the national building stock, in addition to the other assigned heritage categories, see Table 5. Using a scale offered the advantage that the ranked buildings could be flexibly subdivided in subsequent analyses, together with the heritage experts. Ten buildings in Visby, Nässjö, Gothenburg, Uppsala, and Stockholm, respectively, were subsequently subjected to expert quality control during the workshop. All of the inspected cases were confirmed to exhibit features that are currently associated with heritage value, supporting the validity of the approach. However, since the evaluation focused primarily on the buildings with a score of predicted heritage indicator score of 50, there remains a risk that potential visible features representing heritage values in buildings with lower scores were overlooked.

Nationwide prediction of heritage values remains a methodological challenge. Dabrock et al.[38] demonstrated that indications of such values can be approximated using features extracted from Google Street View façade images combined with machine learning, with validation showing reasonable agreement with expert inventories. While the approach yielded promising results in the Stockholm municipality, performance was lower in other regions of Sweden. For the EPBD process, the analysis therefore considered multiple sources of heritage building information. Buildings officially protected at the national level (Byggnadsminne), the Räkna Q dataset documenting buildings locally protected under the Swedish Planning and Building Act (PBA), and age-based classifications (Buildings constructed before 1920, but also buildings with the default construction year 1929 given to buildings built in 1929 or earlier) were first all assigned as buildings with heritage value in the EPBD process. Thereafter, buildings with predicted heritage indicator scores of 50 or more were assigned to the EPBD category Predicted heritage value, see Table 5.

Table 5 Heated floor areas of all buildings in their respective predicted heritage indicator score. 50 was the threshold used to assign heritage value for the EPBD. The first three columns are the official heritage groups identified by the National Heritage Board. The columns sum to 100%.

| Predicted heritage indicator [1-100] | Byggnadsminne | Räkna Q | Heritage value from age | Predicted heritage value** | No assigned heritage value | Total |
|---|---|---|---|---|---|---|
| 1 | 0.400% | 0.177% | 0.245% | 0.000% | 0.756% | 0.701% |
| 5 | 0.015% | 0.061% | 0.118% | 0.000% | 1.606% | 1.450% |
| 10 | 0.966% | 4.497% | 2.084% | 0.000% | 21.442% | 19.464% |
| 20 | 4.476% | 14.096% | 5.250% | 0.000% | 28.164% | 25.906% |
| 30 | 6.537% | 20.782% | 9.741% | 0.000% | 15.157% | 14.674% |
| 40 | 8.284% | 15.767% | 18.735% | 0.000% | 2.035% | 3.428% |
| 50 | 20.599% | 22.575% | 30.751% | 76.597% | 0.000% | 3.406% |
| 60 | 4.494% | 3.016% | 5.861% | 9.201% | 0.000% | 0.575% |
| 65 | 2.228% | 1.440% | 1.237% | 0.524% | 0.000% | 0.131% |
| 70 | 11.171% | 3.468% | 5.322% | 4.826% | 0.000% | 0.554% |
| 75 | 2.012% | 0.489% | 0.553% | 1.503% | 0.000% | 0.077% |
| 80 | 3.154% | 0.306% | 0.417% | 4.931% | 0.000% | 0.108% |
| 85 | 7.951% | 0.004% | 0.554% | 2.407% | 0.000% | 0.123% |
| 90 | 3.597% | 0.045% | 0.050% | 0.008% | 0.000% | 0.032% |
| 95 | 2.499% | 0.000% | 0.017% | 0.004% | 0.000% | 0.020% |
| N/A* | 21.617% | 13.278% | 19.067% | 0.000% | 30.841% | 29.351% |
| **Total [$Mm^2$]** | **3.76** | **8.77** | **32.87** | **5.00** | **439.70** | **490.10** |

*No available street view image

**This is the category defined in the EPBD process, based on the predicted heritage indicator.

In Table 5, not even buildings officially protected at the national level (Byggnadsminne) received a predicted heritage indicator score of 100, and only 6% were assigned 90 or higher. Predicted heritage indicator score of 50 or more is actually a score that few buildings are given. Given the LLM's aversion to extreme values, threshold choice is highly sensitive. Minor perturbations around the cutoff may lead to substantial shifts in inclusion rates.

## 3.3 Representativeness and bias of the results

Using the predicted heritage indicator to assign heritage values to specific buildings, to be used in the EPBD implementation on the building stock level, all over Sweden, raises a number of questions and concerns. The majority of the Räkna Q buildings were given a predicted heritage indicator score lower than 50, as seen in Table 5. The reason is hypothesised to be because the dataset of Räkna Q buildings is so varied. Locally, municipalities often identify heritage values in everyday environments with less specific visible characteristics, modest design, and in modern buildings. For the buildings officially protected at the national level, Byggnadsminnen, the predicted heritage indicator are higher, but they are a group of buildings with potentially clearer visible character-defining elements, which are often, and more traditionally, related to heritage values. Such elements can be architectural ornaments, classical façade design, and features that suggest high age. This indicates a conservative approach in OpenAI's GPT to what can be considered heritage. The Räkna Q buildings are therefore an interesting group of buildings for this paper[47]. To be noted is that the average construction year is more recent for buildings in Räkna Q (1921) compared to Byggnadsminne (1841).

One risk is that OpenAI's GPT bias favours buildings in the Swedish capital or in affluent areas. In Table 6, we compare buildings with a predicted heritage indicator score of 50 or higher in- and outside of Stockholm, Räkna Q buildings are outside of Stockholm. It should be noted that all buildings built earlier than 1920 and with a registered construction year of 1929 were assigned heritage value from age.

Table 6 Comparison of buildings with heritage indicator score more than 50, in and outside of Stockholm.

| | **Observations with predicted heritage indicator score > 50** | | **All observations** | | **Share of predicted heritage indicator score > 50** | |
|---|---|---|---|---|---|---|
| **Construction period** | **Not Stockholm** | **Stockholm** | **Not Stockholm** | **Stockholm** | **Not Stockholm** | **Stockholm** |
| Before 1900 | 1732 | 1168 | 4170 | 1694 | 41.53% | 68.95% |
| 1900-1919 | 1852 | 868 | 5158 | 1331 | 35.9% | 65.2% |
| 1929 | 1567 | 323 | 6324 | 782 | 24.8% | 41.3% |
| 1920-1944 (excl. 1929) | 838 | 282 | 16328 | 3724 | 5.1% | 7.6% |
| 1945-1959 | 149 | 9 | 25322 | 3217 | 0.59% | 0.28% |
| 1960-1974 | 92 | 13 | 29823 | 1749 | 0.31% | 0.74% |
| 1975-1989 | 172 | 13 | 20514 | 1163 | 0.84% | 1.12% |
| 1990 or after | 165 | 23 | 31115 | 2296 | 0.53% | 1.00% |
| **Grand Total** | **6567** | **2699** | **138754** | **15956** | **4.7%** | **16.9%** |

To assess OpenAI's GPT bias for affluence, the buildings were subdivided based on average disposable income in statistical areas of ca 4000 persons (DESO), see Table 7. Buildings above the dashed line were assigned heritage value based on their age. In Tables 7 and 8, the percentages can be compared in a similar fashion to those in Table 6.

Table 7 Comparison of shares of buildings with predicted heritage indicator score more than 50 in affluence groups of average disposable income [KSEK/year.person].

| **Construction period** | **300>** | **300-400** | **400-500** | **500-600** | **600<** |
|---|---|---|---|---|---|
| Before 1900 | 34.6% | 43.3% | 65.7% | 74.6% | 78.0% |
| 1900-1919 | 30.8% | 37.0% | 60.1% | 77.1% | 69.9% |
| 1929 | 18.8% | 27.6% | 43.4% | 57.3% | 47.6% |
| 1920-1944 (excl. 1929) | 4.3% | 5.0% | 8.0% | 22.8% | 31.6% |
| 1945-1959 | 0.6% | 0.5% | 1.0% | 3.0% | 0.0% |
| 1960-1974 | 0.2% | 0.5% | 1.5% | 1.4% | 2.3% |
| 1975-1989 | 0.6% | 1.0% | 1.9% | 0.9% | 0.0% |
| 1990 or after | 0.5% | 0.5% | 0.7% | 2.2% | 4.1% |
| **Grand Total [%]** | **3.2%** | **5.4%** | **17.6%** | **40.9%** | **36.1%** |
| **Grand Total [pc]** | **70642** | **69343** | **11760** | **2238** | **727** |

Other aspects that could influence OpenAI's GPT-predicted heritage indicator are the identified façade materials and construction types. Wooden structures are more common among the Räkna Q buildings, and one hypothesis is that OpenAI's GPT does not predict heritage values for wooden buildings. Such buildings would probably be found in more modest everyday environments. In Table 8, the buildings have been grouped according to the façade material predicted by OpenAI's GPT. Table 8 indicates that buildings with façade materials of wood indeed have received a lower predicted heritage indicator score. In addition to construction material, the prompt also included a feature for construction type. Wooden buildings with a construction type restimmerhus or resvirkeshus, for which heritage value is well established in literature, received a high predicted heritage indicator score.

Table 8 Comparison of shares of buildings with predicted heritage indicator score more than 50 in groups of predicted façade material.

| Construction period | Brick | Metal | Plaster | Stone | Wood |
|---|---|---|---|---|---|
| Before 1900 | 68.7% | 0.0% | 53.3% | 79.7% | 24.8% |
| 1900-1919 | 58.3% | 0.0% | 47.1% | 81.4% | 15.3% |
| 1929 | 43.6% | 0.0% | 28.2% | 79.2% | 10.4% |
| 1920-1944 (excl. 1929) | 8.1% | 0.0% | 5.6% | 48.4% | 3.2% |
| 1945-1959 | 0.5% | 0.0% | 0.4% | 16.7% | 1.0% |
| 1960-1974 | 0.2% | 0.1% | 0.5% | 5.7% | 0.5% |
| 1975-1989 | 0.6% | 0.1% | 2.2% | 14.7% | 0.7% |
| 1990 or after | 0.6% | 0.2% | 0.6% | 6.3% | 0.5% |
| **Grand Total [%]** | **5.7%** | **0.1%** | **8.6%** | **55.6%** | **3.2%** |
| **Grand Total [pc]** | **49792** | **6890** | **58049** | **450** | **35515** |

## 3.4 Application in the EPBD implementation

For the EPBD process, all analyses were done on the building stock scale. For these purposes, the Swedish National Heritage Board was comfortable using the methods described in this article. However, even though the result of the building stock analyses is always a group of buildings, the unit of the analyses is individual buildings. During the workshop with the National Heritage Board heritage experts stated, based on the facade images alone, that some buildings in this group potentially could qualify for official protection. Sweden has 290 municipalities, and each municipality was sent information about its respective building stock, including the groups of buildings with assigned heritage values.

The threshold was intentionally chosen to only include an additional 1-2% of the building stock in the group predicted heritage value, beyond the first three groups, in Table 5. This limit was chosen based on estimations from experts at the National Heritage Board and on the results of the predictions. The threshold point, predicted heritage indicator score 50, is actually a rather high bar to pass. OpenAI's GPT is relatively reluctant to set higher heritage values. Including few additional buildings using the LLM results and observing a randomised sample of those buildings made decision-makers at the National Heritage Board convinced enough to use the results in the EPBD implementation process. The National Heritage Board decided that potentially missing the identification of some buildings was less of a problem for analyses on the building stock level. A result of this exploratory work is to exemplify the importance of tailoring methods for specific purposes of authorities rather than designing methods with general purposes.

# 4 Discussion

The results show a conservative approach to what can be considered to represent heritage values today. Research in the field of heritage management has identified a need to broaden heritage selection and increase the inclusivity in such evaluations in order to represent more varied histories and cultures[10]. Since what we select and value as heritage changes based on present needs and tastes, assessments should not be considered fixed or intrinsic in the buildings themselves, but part of a continuous process[9]. AI is not able to renew or critically analyse a selected group of heritage buildings to make them more inclusive or add new perspectives to heritage management. That is outside the epistemological limitation of AI usage in the article. As we have demonstrated in this article, AI blindly reproduces bias in favour of buildings in urban and affluent areas, and will certainly overlook modern buildings as potential heritage.

This article has demonstrated that it is possible to process complex information for tasks at high speed. However, AI methods have the risk of being "too easy" and of thus being blindly applied as a replacement for traditional approaches for deriving heritage values of buildings to save time and money. While the methods have a high potential for efficiency increases in this field, they should not be applied without the input and oversight of heritage experts and should rather be seen as a support and source of supplementary information. Ideally, the methods can be applied to reduce cumbersome and monotonous work for heritage experts, so heritage experts can focus on strategic and more complex tasks that require accountability.

One key aspect when controlling the quality of predictions is to use parameters rather than text answers from the LLM. The risks of sycophancy of the LLM are mitigated when the model cannot reply in texts. Parameters can then easily be examined by human beings and compared to existing registers and subsets. However, human judgment can be affected by the automated assessment provided by the LLM, a phenomenon known as automation bias[48]. To create a structure for assigning heritage values, checklists are used to justify assessments of building elements that add heritage value in order to achieve uniformity in assessments. It is interesting to make a parallel between this human way of justifying assessments of aspects of heritage value and the automated labelling process presented in this paper. LLMs enable analyses that are adapted to this human way of making assessments, using checklists. This enables both quality control of predictions and the integration of the results in existing human methods and datasets of heritage assessment. Checklists are used for inventories of other parts of the building sector. Environmental inventories, permit processes for renovations, Life Cycle Analyses, and even in real estate taxation and pricing, are potential future applications of multimodal LLM, especially if agent models are further developed[49].

The black-box nature of the LLMs makes it impossible to reconstruct how GPT generates its output[4,6]. While features such as construction year and number of storeys can be evaluated as false or correct, other features, such as heritage value, require more information and context in order to be verified[35]. How LLMs are biased in terms of cultural, social, and religious aspects is hidden and difficult to assess. In this article, we find that there is an overrepresentation of predicted heritage values in affluent areas and buildings in Stockholm. However, richer people live in buildings with high heritage values (see Supplementary Material 2). Buildings in Stockholm are older, and construction styles often first appeared in the Swedish capital[50]. These biases derive from the vast LLM training data produced by humanity, but the LLMs are also built to be sycophantic and to reproduce bias. Keeping this in mind is highly relevant when using the data for any form of legislation that could discriminate against certain individuals or groups of people. Apart from the potential implications of decision-making and legislation based on the output of AI, legal aspects are also relevant to the methodology itself. Aspects such as who has the right to take pictures of private buildings, who owns the images, how the images may be used[37] and what data is used for the training of LLMs[51].

Due to the changing nature of heritage assessments, which are influenced by their purpose and context in time and place, it could be questioned whether it is even possible to have a comprehensive register of buildings with heritage values. Even the effort itself can be questioned. Listing buildings with heritage values for the sake of stipulating protection is being questioned, especially with the growing concern of sustainable management of the built environment. Researchers and practitioners in the heritage field point instead to a need to increase the integration of heritage knowledge in a larger context of sustainable management of the existing building stock, and listings may have the opposite effect[53]. Such selections risk creating a "silo" instead of integration, or a situation where the selected heritage has value, and the rest does not[53]. The potential to apply more flexible approaches exists in current legislation, particularly the requirement of caution, which should be applied to all buildings, although the implementation of this clause requires improvements. Case-by-case decisions are thus being made on the local level, which require competence and integration of preservation concerns in processes of change, but what kind of building data these decisions should be based on is now the subject of reconsideration. Carbon emission from the usage of LLMs is an additional important aspect that needs to be considered when using repetitive prompts for entire building stocks[54,55]. The prompt and model should not use more computational power than is required[56]. This is especially important if reasoning models are used, which can require substantially more tokens per building.

## 4.1 Future applications

The Swedish National Heritage Board has the intention to continue exploring the possibilities with AI methods. Ideas that have been raised are to develop methods for municipalities to use and also for evaluation of national grant schemes. It will be stressed to any user that the heritage features need to be verified. For this reason, we have demonstrated how AI reproduces bias in this article.

If AI is to be used for identifying characteristics in built environments, then AI could be prompted to identify certain visible features in different building categories that are rare or typical for specific building types, in order to provide data for human analyses and revisions of such descriptions. Additionally, the use of AI can increase the speed at which assessments of built environments are made, thus providing opportunities for keeping records updated. By making the process of recording building details more efficient, the conservation officers in municipalities would be able to put more time and effort into working on developing strategies for local heritage management.

The methods could also be used for predicting other features. Predicting the presence of building materials has importance for reuse[57], circularity in the building industry[58], and for the identification of hazardous building materials[59]. Another aspect is the inclusion of other data sources. Satellite images or 3D model images of buildings could be included for a better understanding of the specific buildings in question.

The methods described in this paper are currently being used again in the EPBD implementation process. Next, the street view images are used to predict renovation state and building usage for the Swedish Building Renovation Plan. For buildings that do not have an Energy Performance Certificate and are registered in the property register as 'Other' or "Societal function – unspecified". Building usage is assumed to be less complicated for LLMs to predict than heritage values based on street view images.

The EPBD must be implemented by all European member states. In Germany, the country with the highest final energy consumption of the building stock[60], there is also no national dataset on the heritage value of buildings that reliably represents the entire building stock. Instead, data sources are heterogeneous and fragmented as the heritage legislation differs by federal state and the responsibility for cataloguing heritage buildings lies with the local heritage protection authorities at the municipal level. An automated approach could improve the identification of heritage values when developing strategies for increasing the energy efficiency of the building stock at the national level. The street-view imagery from Google Street View is available in Germany as in Sweden, and a similar procedure for retrieving and processing images can be applied. However, in Germany, no nationwide building database is available, equivalent to the dataset used in this article. Therefore, as a replacement, other building data sources could be used, such as the German dataset ETHOS.BUILDA[61] or International datasets such as OpenStreetMap, EUBUCCO, or Overture Maps. These latter have the advantage of being available for all European countries.

# 5 Conclusion

This paper has demonstrated that zero-shot predictions by multimodal LLMs can have value for understanding the building stock and in decision-making, specifically in the EPBD implementation process in Sweden. However, the methods need to be used carefully, there are a number of pitfalls and lessons to learn. The authorities' checklists for initial ocular surveys for heritage assessment were used to design the prompt. This facilitated understanding of GPT errors, supported legitimacy and reduced risks of sycophancy, compared to using sentences produced by GPT. When the heritage values were assigned finally by authorities using the predicted heritage indicator score, only an additional 1-2% of buildings with the highest score were included.

In this paper, ethical dimensions in terms of socio-economic biases were analysed by comparing predicted heritage values for buildings subdivided into inhabitant income groups. We found that buildings where richer people live tend to receive a higher predicted heritage indicator score. OpenAI's GPT also gives higher predicted heritage indicators to buildings in Stockholm when controlling for building age. Richer people tend to live in buildings with heritage values and building techniques were often first introduced in Stockholm, but the point is that LLM reproduces bias. When using zero-shot predictions by multimodal LLMs, there are clear efficiency gains, but the generated results cannot be used without critical oversight from building heritage experts with an awareness of bias. Defining and renewing the way in which heritage values are assigned and used in the development of the built environment requires heritage expertise, while AI models can do the more menial, repetitive tasks of gathering data necessary in the process.

# Acknowledgments

The authors would like to thank Ebba Gillbrand, Camilla Altahr-Cederberg and Therese Sonehag at the Swedish National Heritage Board for collaboration, discussions and workshop participation.

# Author Contribution

T.J.: Conceptualisation, data processing, writing – original draft. M.M.: Conceptualisation, data analysis, writing – original draft. K.D.: Writing, data analysis – review and editing. A.D.: Writing, workshop management – review and editing. I.C.-R. Writing – review and editing. All authors reviewed the manuscript.

# Declaration of interest statement

No potential conflict of interest was reported by the author(s).

# Data and source code availability statement

Swedish register data are protected under confidentiality agreements. However, sharing of data is possible for research purposes in joint projects. The building image data is available from GoogleStreetView. A module enabling testing the methods and the source code is available at GitHub.

# Declaration of generative AI and AI-assisted technologies in the writing process

During the preparation of this work, the authors used DeepL for linguistic revisions. After using this tool, the authors reviewed and edited the content as needed and take full responsibility for the content of the published article.

# Funding

This work was funded by the Swedish National Heritage Board research fund: Dnr RAÄ-2024-1995.